\documentclass[aps,prl,twocolumn,superscriptaddress,amsfonts,showpacs]{revtex4}

\usepackage{amsmath}
\usepackage{graphicx,color}
\usepackage{bm}
\usepackage{natbib}
\begin{document}

\title{Tunneling anisotropic magnetoresistance
in Fe/GaAs/Au junctions: orbital effects}

\author{M.~Wimmer}
\affiliation{Institut f\"ur Theoretische Physik, Universit\"{a}t Regensburg, 93040 Regensburg, Germany}
\author{M.~Lobenhofer}
\author{J.~Moser}
\affiliation{Institut f\"{u}r Experimentelle und Angewandte Physik, Universit\"{a}t Regensburg, 93040 Regensburg, Germany}
\author{A.~Matos-Abiague}
\affiliation{Institut f\"ur Theoretische Physik, Universit\"{a}t Regensburg, 93040 Regensburg, Germany}
\author{D.~Schuh}
\author{W.~Wegscheider}
\affiliation{Institut f\"{u}r Experimentelle und Angewandte Physik, Universit\"{a}t Regensburg, 93040 Regensburg, Germany}
\author{J.~Fabian}
\author{K.~Richter}
\affiliation{Institut f\"ur Theoretische Physik, Universit\"{a}t Regensburg, 93040 Regensburg, Germany}
\author{D.~Weiss}
\affiliation{Institut f\"{u}r Experimentelle und Angewandte Physik, Universit\"{a}t Regensburg, 93040 Regensburg, Germany}

\date{\today}

\begin{abstract}
We report experiments on epitaxially grown Fe/GaAs/Au tunnel junctions
demonstrating that the tunneling anisotropic magnetoresistance (TAMR) effect
can be controlled by a magnetic field. Theoretical modelling shows that the
interplay of the orbital effects of a magnetic field and the
Dresselhaus spin-orbit coupling in the GaAs barrier leads to an independent
contribution to the TAMR effect with uniaxial symmetry, whereas the 
Bychkov-Rashba spin-orbit coupling does not play a role. The effect is
intrinsic to barriers with bulk inversion asymmetry.
%We report experiments on epitaxially grown Fe/GaAs/Au tunnel junctions
%showing a peculiar magnetic field dependence of the tunneling anisotropic
%magnetoresistance (TAMR) effect. Theoretical modelling shows that the
%interplay of the orbital effects of a magnetic field and the
%Dresselhaus spin-orbit coupling in the GaAs barrier leads to an independent
%contribution to the TAMR effect with uniaxial symmetry, intrinsic
%to semiconductor barrier with bulk inversion asymmetry.
\end{abstract}


\maketitle

Magnetic tunnel junctions (MTJs) are prominent examples of spintronic
devices \cite{Zutic2004,Fabian2007} and have reached already
technological importance \cite{Chappert2007}. Typically, the
resistance of a MTJ depends on the relative
orientation of two ferromagnetic layers
\cite{Zutic2004,Fabian2007}. Hence it came as a surprise when
experiments on MTJs with only one ferromagnetic GaMnAs
layer showed a sizeable spin valve effect \cite{Gould2004}. Since then,
this tunneling anisotropic magnetoresistance (TAMR) effect has
been observed in tunnel junctions involving various materials
\cite{Ruster2005,Saito2005,Moser2007,Park2008,Ciorga2007} as well as
nanoconstrictions and break junctions \cite{Giddings2005,
Bolotin2006,Ciorga2007}. Amongst these experiments, the TAMR effect
in Fe/GaAs/Au MTJs \cite{Moser2007} stands out due to its
qualitatively different origin: Whereas
the TAMR effect usually originates from properties of the magnetic
layer, namely a spin-orbit induced anisotropic density of states
\cite{Gould2004,Ruster2005,Saito2005, Park2008, Giddings2005} in the
ferromagnet or surface states \cite{Chantis2007,Burton2007,Khan2008}, 
the TAMR in the Fe/GaAs/Au MTJ was attributed to an interference of 
Bychkov-Rashba spin-orbit
coupling (SOC) at the barrier interface and the Dresselhaus SOC
inside the barrier, i.e.~to properties of the tunneling process
itself. Moreover, the size and sign of the effect in this MTJ 
can be tuned by the bias voltage.

In this Letter, we show experimentally that the TAMR in
Fe/GaAs/Au MTJs can also be controlled by a
\emph{magnetic} field. Our theoretical calculations ascribe this effect to 
an interplay of the orbital effects of the magnetic field and the 
Dresselhaus SOC in the GaAs barrier. This interplay leads to an
independent TAMR contribution with uniaxial symmetry 
and is intrinsic to semiconductor barriers with bulk inversion asymmetry. 
Whereas spin-orbit effects are usually controlled through the electric field
dependence of the Bychkov-Rashba SOC \cite{Zutic2004,Fabian2007}
(bias voltage in the case of the TAMR \cite{Moser2007}), the magnetic
field dependence of the TAMR is only linked to the 
Dresselhaus SOC; interestingly the Bychkov-Rashba SOC does not play 
a role here. As we show below, this is due to the different symmetries of the SOCs.
Furthermore, in our analysis
we find it important to include the orbital effects of the magnetic field
in both the kinetic and SOC terms of the Hamiltonian, as both terms
give rise to large competing contributions, resulting
in a net TAMR effect in good agreement with experiment. 

\begin{figure}
\includegraphics[width=0.98\linewidth]{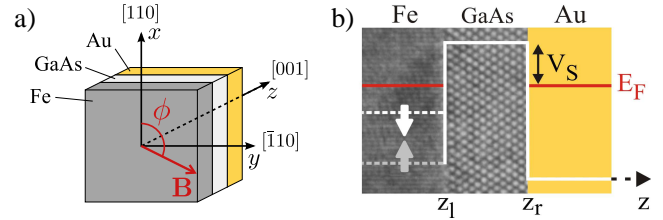}
\caption{(Color online) (a) Sketch of the Fe/GaAs/Au MTJ.
(b) Schematic of the conduction band profile. The grey
background is a transmission electron micrograph of an epitaxial Fe/GaAs
interface displaying the $8\,$nm thick GaAs barrier.}
\label{fig1}
\end{figure}

The type of tunneling device studied here is sketched in
Fig.~\ref{fig1}. We explored 8 different samples all showing
the same orbital effects discussed below. We hence focus here on one
sample which consists of a $13\,$nm thick Fe layer, grown epitaxially
on a $8\,$nm thick GaAs-tunneling barrier, and a Au top electrode
\cite{Moser2007}. The GaAs barrier was grown by molecular beam
epitaxy on sacrificial AlGaAs layers and capped with As to prevent
oxidation during transport to a UHV magnetron sputtering system.
There the As cap was removed at $T=250^\circ$C and Fe was grown at
room temperature. Epitaxial growth of the Fe film was monitored by
in-situ RHEED. The Fe-layer is finally covered with $50\,$nm Co, and
$150\,$nm Au and serves as back contact. To prepare the top Au-contact
on the other side of the GaAs tunnel barrier, the wafer is glued
upside down to another substrate and the original substrate is
etched away. By employing optical lithography, selective etching and
UHV-magnetron sputtering a circular, $13\,\mu$m wide and $100\,$nm thick
Au contact is fabricated.

The measurements were carried out at a temperature of $4.2\,$K inside a
variable temperature insert of a $^{4}$He-cryostat. The device was
placed in a rotatable sample holder allowing a 360$^\circ$
in-plane rotation in the magnetic field $\mathbf{B}$
of a superconducting solenoid. The direction of $\mathbf{B}$ with
respect to the hard-axis of the Fe layer in [110]-direction
(nomenclature with respect to GaAs crystallographic directions) is
given by the angle $\phi$ (Fig.~\ref{fig1}). The resistance drop
across the tunnel barrier was measured in
four-point-configuration using a HP 4155A semiconductor parameter
analyzer with the Au-contact grounded.

\begin{figure}
\includegraphics[width=\linewidth]{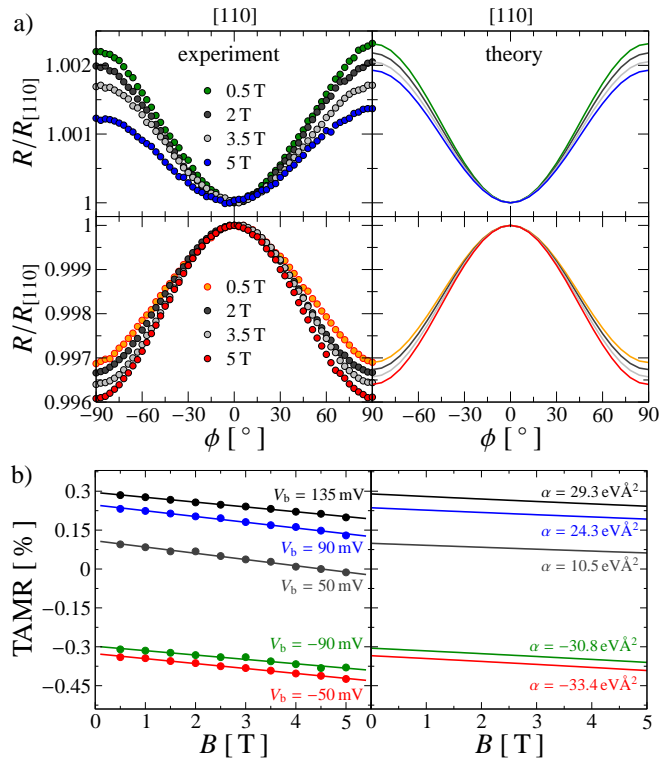}
\caption{(Color online) Comparison of experimental results (left panels) and
numerical simulations (right panels) for the TAMR.
(a) Angular dependence of $R(\phi)$ at various magnetic fields
for an applied bias voltage of $V_\text{b}=90\,$mV (upper panels) and
$V_\text{b}=-90\,$mV (lower panels).
(b) $B$-dependence of the TAMR ratio \eqref{TAMR} for different bias voltages.
Lines in the left panel are a linear fit to the experimental
data, shown as dots. 
%The experiments were carried out at $4.2\,$K. The
%value of $\alpha$ used in the numerics
%was obtained from a fit to the experimental data at $0.5\,$T.
}\label{fig2}
\end{figure}

To measure the TAMR we rotated the sample by 180$^\circ$ in a
constant external magnetic field. The magnetic field strength was
always high enough to align the magnetization $\mathbf{M}$ along
$\mathbf{B}$. Fig.~\ref{fig2}(a) shows the results of such
$\phi$-scans for various values of the magnetic field between 0.5 T
and 5 T and the two bias voltages,  $+90\,$mV (upper left panel) 
and $-90\,$mV (lower left panel). 
The TAMR $R(\phi)/R_{[110]}$ shows the distinct uniaxial 
anisotropy characteristic for this
system \cite{Moser2007}. As demonstrated recently, the TAMR strongly
depends on the applied bias voltage and is connected to a bias
dependent sign and strength of the Bychkov-Rashba parameter
\cite{Moser2007}. For $\mathbf{M}
\parallel [110]$ we always get a resistance maximum for $+90\,$mV,
but a minimum for $-90\,$mV. This behavior is in accord
with the one observed by Moser \emph{et al.}~\cite{Moser2007} and
occurs for all samples investigated. In the simplest model the TAMR
$R(\phi)/R_{[110]}-1\sim \alpha \gamma (\cos(2\phi)-1)$ where 
$\alpha$ and $\gamma$ are Bychkov-Rashba and Dresselhaus parameters. While
$\gamma$ is a material parameter, $\alpha$ is
obtained by fitting the angular dependence $R(\phi)/R_{[110]}$ (see
below).

With increasing magnetic field strength both the traces for positive and
negative bias voltages are bent towards lower resistance values. If we
define the TAMR ratio as
\begin{equation}\label{TAMR}
    \text{TAMR}=\frac{R_{[\bar 110]}-R_{[110]}}{R_{[110]}},
\end{equation}
in which $R_{[\bar 110]}$ is the resistance for $\phi=+90^\circ$,
the magnitude of the TAMR ratio decreases for positive bias voltages but
increases for negative ones. This TAMR value measured as function of $B$
is displayed in the left panel of Fig.~\ref{fig2}(b) 
for magnetic field strengths up to 5 T. Note that the TAMR 
vanishes for a bias voltage of $+50\,$mV
at about 4.5 T but reappears again upon 
further increasing $B$. The magnetic field dependence of the TAMR ratio is
in all cases linear. The slope $\Delta\text{TAMR}/\Delta B$ of the
best-fit line is nearly the same for all bias voltages indicating
that the $B$-dependence of the TAMR is independent of the
applied voltage. The experimental data in Fig.~\ref{fig2} are
compared to model calculations discussed below.

The importance of orbital effects for charge tunneling
has been pointed out already in the literature
\cite{Eaves1986}. Here, we focus on orbital effects on \emph{spin-dependent} 
tunneling. In order to explain the
experimental findings, we employ the spin-orbit based model for the
TAMR effect of Refs.~\cite{Moser2007,Fabian2007,MatosAbiague2009} and include the
orbital effects of the magnetic field. We choose the coordinate
system such that the $x$, $y$, and $z$-directions are along the
$[110]$, $[\bar{1}00]$, and $[001]$ crystallographic directions, and
consider an in-plane magnetic field $\mathbf{B}=B \mathbf{n}$,
where $\mathbf{n}=(\cos \phi, \sin \phi,0)$ is a unit vector 
forming an angle $\phi$ with the $x$-axis (see Fig.~\ref{fig1}(a)). 
The Hamiltonian is given as
$H=H_0+H_\text{BR}+H_\text{D}$, where
\begin{equation}\label{eq2}
H_0=+\frac{1}{2}\, \boldsymbol{\pi} \frac{1}{m^*(z)} \boldsymbol{\pi}
+V(z)+\frac{\Delta(z)}{2}\,\mathbf{n}\cdot\boldsymbol{\sigma}\,.
\end{equation}
Here, $\boldsymbol{\pi}=-i\hbar\boldsymbol{\nabla}+e\mathbf{A}$, where $\mathbf{A}$ is the
magnetic vector potential and $-e$ the electron charge.
$m^*(z)$ is a position-dependent
effective mass with $m^*(z)=0.067m_\text{e}$ in the GaAs barrier
and $m^*(z)=m_\text{e}$ in the Fe and Au layer, where $m_\text{e}$
denotes the bare electron mass. $V(z)$ is the conduction band profile in growth
direction $z$. The GaAs Schottky barrier height is
given by $V_\text{S}=0.75\,$eV. 
The ferromagnetism in the Fe layer is described in
terms of a Stoner model \cite{Stoner1939b} with spin splitting
$\Delta(z)$. $\Delta(z)$ and $V(z)$ are chosen such that
the Fermi wavevector in Fe is $k_{\text{F,Fe}}^\uparrow = 1.05 \times 10^{-10}
m^{-1}$ and $k_{\text{F,Fe}}^\downarrow = 0.44 \times 10^{-10}m^{-1}$
for majority and minority electrons \cite{Wijn1991}, respectively, and in Au
$k_{\text{F,Au}} = 1.2 \times 10^{-10}m^{-1}$ \cite{Ashcroft1988}. The
Zeeman splitting in GaAs and Au is much smaller than any relevant
energy scale in the system and can be neglected, as is also confirmed
by numerical simulations.

The SOC due to the structural inversion asymmetry (SIA)
at the Fe/GaAs-interface can be written as \cite{Bychkov1984}
\begin{equation}\label{eq:rashba}
H_\text{BR}=\frac{\alpha}{\hbar} (\sigma_x \pi_y - \sigma_y \pi_x) \delta(z-z_l)\,,
\end{equation}
where $z_l$ denotes the position of the Fe/GaAs-interface. As in
Refs.~\cite{Moser2007,Fabian2007,MatosAbiague2009} we use the Bychkov-Rashba parameter
$\alpha$ as a fitting parameter to reproduce the bias dependence of the
TAMR effect; $\alpha=\alpha(V_\text{b})$ \cite{Gmitra2009}.

Finally, the SOC due to the bulk inversion asymmetry (BIA) of the zinc-blende
GaAs barrier takes the form \cite{Dresselhaus1955}
\begin{equation}\label{eq3}
H_\text{D}=-\frac{1}{\hbar}(\sigma_x \pi_y + \sigma_y \pi_x) \frac{\partial}{\partial z}
\gamma(z) \frac{\partial}{\partial z}\,,
\end{equation}
where the bulk Dresselhaus parameter $\gamma=24\,$eV\AA$^3$ in the
GaAs barrier and $\gamma=0$ elsewhere. Note that the orbital effects of $B$
are also included in the SOC terms.

With the gauge $\mathbf{A}(z)=
(B \sin(\phi) z, -B \cos(\phi) z, 0)$ 
the Hamiltonian $H$ is translationally invariant in
$x$ and $y$-direction, and the in-plane wave vector $\mathbf{k}_{||}=
(k_x,k_y,0)$ is a good quantum number. The conductance in the
Landauer-B\"uttiker formalism \cite{Datta2002} is then given as
$G=\frac{e^2 S}{h (2\pi)^2}\int d\mathbf{k}_{||} T(\mathbf{k}_{||})$,
where $S$ is the cross-sectional area of the junction, and $T(\mathbf{k}_{||})$
is the total transmission probability (including different spin species)
for the transverse wave vector $\mathbf{k}_{||}$ at the Fermi energy
$E_\text{F}$. We calculate $T(\mathbf{k}_{||})$ from the
scattering wave functions \cite{caveatA}; those are obtained numerically from
a tight-binding approximation to $H$, using the
method of finite differences on a one-dimensional grid with lattice
spacing $a=0.01\,$nm \cite{delta}
and the recursive Greens function technique
\cite{Wimmer2009}.

In Fig.~\ref{fig2} we compare the results of the numerical simulations on the
$B$-dependence of the TAMR with the corresponding
experimental data. For this, we fit the parameter
$\alpha$ at $B=0.5\,$T for every value of the bias voltage
$V_\text{b}$ to the experimental data. The dependence on $B$ can then 
be calculated without fitting any further parameter.

Figure \ref{fig2}(a) shows the angular dependence of the
TAMR effect for different values of the bias voltage and magnetic
field. The numerical simulations show the same trend as the experiment:
The magnitude of the TAMR effect decreases with increasing $B$,
when the effect is positive, and it increases, when the effect is
negative. Furthermore, the numerical calculations reproduce
the experimentally found change with magnetic field within a factor
of $1.5-2$. This is an especially satisfying agreement, given the
fact that the $B$-dependence is calculated without
any fitting parameter.
The numerically calculated magnetic field dependence of the TAMR
ratio is shown in Fig.~\ref{fig2}(b). As the experiment, we find
a linear dependence on $B$, with a slope
that is nearly independent of $\alpha$, i.e.~the bias voltage.
Again, the numerics underestimates the slope only by a
small factor of $1.5-2$.

Having established that our model is able to reproduce
both qualitatively and quantitatively the experimental findings, we
now develop a phenomenological model to highlight
the underlying physics. 
In Refs.~\cite{Moser2007,Fabian2007,MatosAbiague2009}
it was shown that in the absence of a magnetic field, $T(\mathbf{k}_{||})$ 
can be expanded in powers of the SOC in the form
$T(\mathbf{k}_{||})=T^{(0)}(k_{||})+T^{(1)}(k_{||})\,\mathbf{n}
\cdot\mathbf{w}(\mathbf{k}_{||}) + T^{(2)}(k_{||})\,(\mathbf{n}
\cdot\mathbf{w}(\mathbf{k}_{||}))^2 +\dots$, where the $T^{(n)}(k_{||})$
are expansion coefficients and $\mathbf{w}(\mathbf{k}_{||})=
((\tilde{\alpha}-\tilde{\gamma}) k_y, -(\tilde{\alpha}+\tilde{\gamma}) k_x,0)$ the
effective spin-orbit field obtained by averaging the
spin-orbit field $\mathbf{B_\text{SO}}(z)$, $H_\text{D}+H_\text{BR}=\mathbf{B}_\text{SO}(z)\cdot
\boldsymbol{\sigma}$, over the unperturbed states of the system. The effective
spin-orbit parameters are given by $\tilde{\alpha}=\alpha f_\alpha(k_{||})$
and $\tilde{\gamma}=\gamma f_\gamma(k_{||})$.
To second order in the SOC, the conductance was then found as
\begin{equation}\label{eq:g_noB}
G(\phi)=G_0 + g^{(2)} \alpha\gamma \cos(2\phi)\,,
\end{equation}
where $G_0$ is the
angular-independent part of the conductance and $g^{(2)}$ a coefficient that is
independent of the spin orbit parameters (for details see 
Refs.~\cite{Fabian2007,MatosAbiague2009}).

In the presence of a magnetic field, the transmission can still be expanded in
powers of the SOC, albeit with $B$-dependent coefficients 
$T_B^{(n)}(\mathbf{k}_{||})$ and spin-orbit field 
$\mathbf{w}_B(\mathbf{k}_{||})$. Below, we derive approximate relations 
for $T^{(n)}_B(\mathbf{k}_{||})$ and $\mathbf{w}_B(\mathbf{k}_{||})$, 
valid to linear order in $B$, in terms of their counterparts at 
$B=0$, $T^{(n)}(k_{||})$ and $\mathbf{w}(\mathbf{k}_{||})$.

First, we consider the orbital effects of $B$ on the kinetic energy
term of the Hamiltonian. The kinetic energy associated with
$\mathbf{k}_{||}$ increases the effective barrier
height, and hence $T(\mathbf{k}_{||})$ is
sharply peaked at $\mathbf{k}_{||}=0$ in the absence of a magnetic
field. For $B\neq 0$ however, the effective barrier height is smallest
for an in-plane wave vector $\mathbf{k}_{||,0}$ with
$\langle (k_{x,0}+eB/\hbar \sin(\phi) z)^2\rangle =0$ and
$\langle (k_{y,0}-eB/\hbar \cos(\phi) z)^2\rangle =0$, where
$\langle \dots \rangle$ denotes a quantum mechanical average. Thus,
the maximum of the transmission is shifted to
$\mathbf{k}_{||,0}=(-b_1 B \sin(\phi), b_1 B \cos(\phi),0)$
where $b_1$ depends on $\langle z\rangle$ and $\langle z^2\rangle$, and hence
we assume $T^{(n)}_B(\mathbf{k}_{||})\approx
T^{(n)}(\sqrt{(k_x-k_{x,0})^2+(k_y-k_{y,0})^2})$. This shift
can be interpreted as an effect of the Lorentz force. In addition
to the shift of the maximum, the overall transmission decreases
\cite{Eaves1986}. However, this decrease is quadratic in $B$ and will consequently
be neglected. Apart from $T_B(\mathbf{k}_{||})$, also the effective
spin-orbit field is shifted in momentum space, $\mathbf{w}_B(\mathbf{k}_{||})\approx
\mathbf{w}(k_x+b_2 B \sin(\phi),k_y-b_2 B \cos(\phi))$, where $b_2$ is
a constant that depends on $\langle z\rangle$ only, as the SOC terms are
linear in momentum. Therefore we can in general expect $b_1\neq b_2$.

\begin{figure}
\includegraphics[width=\linewidth]{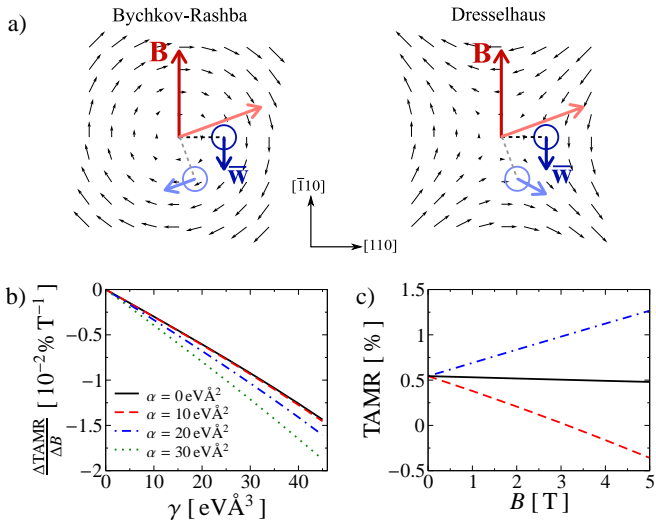}
\caption{(Color online) (a) Schematic picture of the influence of $\mathbf{B}$
  on the TAMR: $\mathbf{B}$
  and $\bar{\mathbf{w}}=\int d\mathbf{k}_{||} T^{(1)}_B \mathbf{w}_B$
  are shown in relation to the effective
  Bychkov-Rashba and Dresselhaus spin-orbit fields. The shift of
  the transmission maximum is indicated by a blue
  circle; the situation for two different angles $\phi$ is shown in
  dark and light color. (b) Slope of
  the $B$-dependence of the TAMR effect, $\Delta
  \text{TAMR}/\Delta B$, as a function of the Dresselhaus parameter
  $\gamma$ for various values of $\alpha$. (c) $B$-dependence of the TAMR
  effect when $\mathbf{A}$ is only included in the
  kinetic term (red dashed line), only in the spin-orbit term (blue
  dash-dotted line), or in both terms (black solid line). }\label{fig3}
\end{figure}

With these approximations we can now obtain the magnetic field
corrections to the conductance of Eq.~\eqref{eq:g_noB} by evaluating
$\int d\mathbf{k}_{||} T(\mathbf{k}_{||})$ in orders of the SOC.
The zeroth order term remains unchanged upon integration, and the corrections to
the second order term are quadratic in $B$, thus being neglected. In contrast,
the first order term that vanishes in the absence of a magnetic field
\cite{Moser2007,Fabian2007,MatosAbiague2009} gives a contribution linear in $B$:
\begin{equation}
\begin{split}
\frac{e^2 S}{h (2\pi)^2}\,\mathbf{n}\cdot \int d\mathbf{k}_{||} T^{(1)}_B(\mathbf{k}_{||})
\mathbf{w}_B(\mathbf{k}_{||}) =\\
g_\alpha^{(1)}\alpha B - g_\gamma^{(1)} \gamma B \cos(2\phi)\,,
\end{split}
\end{equation}
where we used the approximations of the previous paragraph
and the fact that terms linear in $\mathbf{k}_{||}$ vanish upon
integration \cite{Moser2007,Fabian2007,MatosAbiague2009}. The coefficients
$g_{\alpha,\gamma}^{(1)}=\frac{e^2 S}{h (2\pi)^2}(b_1-b_2) \int d\mathbf{k}_{||}
T^{(1)}(k_{||}) f_{\alpha,\gamma}(k_{||})$ do not depend on the
spin orbit parameters. We find a different angular
dependence for the Bychkov-Rashba and Dresselhaus SOC due
to different symmetries of the spin-orbit fields, as shown
in Fig.~\ref{fig3}(a): The Bychkov-Rashba field exhibits
rotational symmetry leading to an angular-independent
contribution, whereas the interplay of $B$ and the Dresselhaus field
leads to an angular dependence with uniaxial symmetry.

The total conductance in a magnetic field is then
\begin{equation}
G(\phi,B)=G_0 + g_\alpha^{(1)}\, \alpha B + (g^{(2)}\, \alpha \gamma - g_\gamma^{(1)}\,
\gamma B) \cos(2\phi)
\end{equation}
valid up to second order in the SOC. The magnetic field dependence of the TAMR ratio
is then
\begin{equation}\label{eq:tamr_model}
\text{TAMR}\propto g^{(2)}\, \alpha \gamma - g_\gamma^{(1)}\,\gamma B\,,
\end{equation}
where we can deduce from the numerical results that the coefficients
$g^{(2)}, g_\gamma^{(1)}>0$. Eq.~\eqref{eq:tamr_model} reproduces all
the characteristic features of the
TAMR observed in experiment: A linear $B$-dependence,
with a bias ($\alpha$)-independent slope. Note that the interplay of
Dresselhaus SOC in the barrier and the orbital effect of the magnetic
field leads to an independent contribution to the TAMR effect which
turns out to have the same uniaxial symmetry as the TAMR effect in the
absence of $B$.

Finally, we verify some aspects of the phenomenological model
by comparing to numerical simulations. In Fig.~\ref{fig3}(b) we show the slope
$\frac{\Delta TAMR}{\Delta B}$ as a function of the Dresselhaus
parameter $\gamma$ that is predicted to be linear in $\gamma$ and independent
of $\alpha$ (Eq.~\eqref{eq:tamr_model}). Indeed, we find a nearly linear dependence
on $\gamma$ and only a weak dependence on $\alpha$, presumably originating
from higher orders in the SOC expansion.
Furthermore, the coefficient $g_\gamma^{(1)}$ in Eq.~\eqref{eq:tamr_model}
depends on $(b_1-b_2)$, i.e.~opposing contributions from the kinetic
and the SOC term. Fig.~\ref{fig3}(c) shows the results of
simulations where the magnetic vector potential is included only in the kinetic term
(dashed line), only in the SOC term (dashed-dotted line), and in both
(solid line). When the magnetic field is included in one term only, we find
large TAMR effects with opposite sign that nearly cancel
in the full Hamiltonian, yielding the small signal observed in experiment
and in the numerics. 

In summary, our experiments and theoretical considerations indicate that the interplay
of the orbital effects of a magnetic field and the Dresselhaus SOC in a
tunnel barrier leads to a contribution to the TAMR effect with uniaxial symmetry.
This effect is predicted to an intrinsic feature of semiconductor barriers with
BIA and not limited to the studied Fe/GaAs/Au tunnel junction.

We gratefully acknowledge financial support by the German Science 
Foundation via SFB 689.


\begin{thebibliography}{24}
\bibitem{caveatA}
  The gauge $\mathbf{A}(z)$ does not
  allow for plane wave solutions in the contacts. Hence, we
  switch off $B$ away from the barrier. Since the
  cyclotron radius in Fe and Au is much larger than any relevant length 
  scale ($l_\text{cycl}\approx 1.5\,\mu$m 
  at $5\,$T), the exact position of this switching is not important.
\bibitem{delta}
  $a$ is chosen very small to accurately approximate the
  $\delta$-function of Eq.~\eqref{eq:rashba} by a rectangular function
  with unit area and width $2a$, centered at the Fe/GaAs-interface.
\end{thebibliography}
\end{document}